# Intelligence as Useful Structure: A Thermodynamic Perspective

Boris Djordjevic
*199 Biotechnologies (SG) Pte Ltd*
boris@199biotechnologies.com

***Abstract*—Thermodynamic accounts of intelligence have advanced rapidly, but they do not yet resolve a prior definitional problem: what makes internal structure intelligent rather than merely ordered? This paper argues that any thermodynamic intelligence metric should include a usefulness filter. The numerator must track internal structure that is relevant to viability, prediction, and control, not merely raw information or benchmark score. Intelligence is therefore defined as the resource-bounded capacity of an adaptive system to acquire, compress, and deploy useful internal structure so as to improve prediction, control, and robustness across a range of environments. Capability and efficiency are separated. Capability is adaptive reach, $A$, defined over a reference distribution of environments or task families. Efficiency is acquisition efficiency, $I_{\text{eff}} = \Delta\frac{C_u}{W_{\text{diss}}}$, where $C_u$ denotes useful internal structure and $W_{\text{diss}}$ physically dissipated work. Concretely, $C_u$ is the predictive and control-relevant information that a system's internal state carries about viability-relevant environmental structure, making it a semantic rather than purely syntactic information term. $I_{\text{eff}}$ is a learning or adaptation metric, not a complete scalar measure of deployed intelligence: trained systems at steady state may have $\Delta C_u = 0$ while still exhibiting high adaptive reach. A toy numerical example, explicit boundary conditions, temperature-normalised reporting conventions, and falsifiable comparisons against size-based and energy-only baselines are provided.**

## I. Introduction

Definitions of intelligence remain fractured across psychology, artificial intelligence, neuroscience, and biology. Behavioural definitions capture success across environments. Thermodynamic work explains the physical costs of information processing and control. Neuroscientific and cybernetic traditions emphasise internal models, prediction, and action. Each strand captures part of the phenomenon, but no single strand by itself cleanly distinguishes intelligent organisation from generic self-organisation.

That distinction is the central problem. Crystals, hurricanes, flames, and driven chemical systems can all build or maintain local order. Some can even retain information about their surroundings. Yet most such systems are not intelligent in any ordinary or scientific sense. What is missing is not merely complexity or causal impact on the world. What is missing is internal structure that matters for the system's own adaptive coupling to the world: structure that helps it predict relevant regularities, guide action and preserve viable futures under perturbation.

The claim is therefore narrow and strong. Thermodynamic metrics of learning and task performance are valuable, but they are not by themselves definitions of intelligence. A definition of intelligence requires an explicit criterion of *useful* internal structure. Such a criterion, situated relative to adjacent formalisms and offered at a level that can be sharpened or falsified, is developed below.

## II. Why a new definition is needed

Four conditions distinguish a workable cross-domain definition of intelligence. It is non-anthropocentric, applying to organisms, artificial agents, and collective systems. It is physically grounded: intelligence is never exempt from finite time, energy, memory, or dissipation. It is selective enough to exclude generic order-production. And it is empirically constraining enough to support measurement and disconfirmation.

Existing definitions solve different parts of this problem. Legg and Hutter define intelligence in terms of success across environments [1]. Chollet reframes intelligence as skill-acquisition efficiency and generalisation under priors and experience constraints [2]. Gignac and Szodorai argue that human and AI research continue to use mismatched conceptual vocabularies [3]. Sternberg likewise argues that intelligence is better understood as successful adaptation than as a single hidden essence [4]. These are important advances, but they remain largely implementation-neutral.

The missing piece is not thermodynamics as such. The missing piece is a principled account of which internal distinctions should count as intelligence-relevant once thermodynamic cost is introduced.

## III. Intellectual background and positioning

### A. Behavioural and psychometric approaches

Legg and Hutter define machine intelligence as the ability to achieve goals across a wide range of environments [1]. Chollet reframes intelligence as skill-acquisition efficiency, with strong emphasis on generalisation rather than benchmark memorisation [2]. These accounts capture adaptive success, but they do not specify what kind of internal organisation makes that success intelligent rather than accidental or physically extravagant.

### B. *Internal models, prediction, and action*

Friston's free-energy framework treats adaptive systems as systems that use generative models to guide perception and action [5]. Ramstead and colleagues extend this line by treating some self-organising systems as encoding beliefs about external states [6]. Pezzulo, Parr, and Friston argue that intelligent behaviour depends on internal models that guide action, and that passive generative AI lacks the closed-loop structure needed for genuine sense-making [7], [8]. These approaches are close in spirit because they connect representation, viability, and control.

### C. *Information, thermodynamics, and adaptive order*

Landauer established that physically realised information processing has irreducible energetic cost [9]. Still and colleagues showed that nonpredictive memory is thermodynamically wasteful: memory that does not improve prediction increases dissipation without improving future coupling to the environment [10]. England's work on dissipative adaptation explains how driven open systems can self-organise to absorb work more effectively [11]. Ali and colleagues show that predictive-coding-like organisation can emerge in recurrent networks trained under energetic pressure [12]. This body of work shows why thermodynamics matters, but it does not by itself specify what makes one learned distinction relevant and another irrelevant.

### D. *Semantic information and selective compression*

That selectivity problem is central in both semantic information theory and the information bottleneck. Kolchinsky and Wolpert define semantic information as the syntactic information a system has about its environment that is causally necessary for maintaining its own existence [13]. Tishby, Pereira, and Bialek formalise the problem of retaining task-relevant information while compressing everything else [14]. These frameworks matter here because they provide the missing filter: the relevant internal structure is not all encoded structure, but structure that matters for viability, prediction, or control.

### E. *Agency, influence, and future options*

Empowerment measures the influence an agent can exert over its future sensor states and therefore supplies a domain-general notion of optionality and control [15]. Tiomkin and colleagues [16] and Kiefer [17] likewise connect action, intrinsic motivation, and entropy-related objectives in ways that bear directly on intelligent control.

### F. *Recent thermodynamic formalisms*

Recent work has pushed thermodynamic intelligence formalisms much further. Takahashi and Hayashi propose two complementary bits-per-joule metrics: thermodynamic epiplexity per joule, which measures learned structural information per unit energy, and empowerment per joule, which measures control capacity per unit energy [18]. Fagan's Conservation-Congruent Encoding framework generalises Landauer-style reasoning to arbitrary conserved quantities and defines intelligence in terms of causal or goal-directed work per unit irreversible information [19]. Perrier introduces a watts-per-intelligence ratio built on an inherited behavioural intelligence score and algorithmic thermodynamic lower bounds [20]. Xu and Li propose Derivation Entropy, connecting Shannon entropy, logical depth, and thermodynamic cost [21]. Han introduces intelligence inertia, a nonlinear cost of reconfiguration [22]. These are substantive advances on which the argument below depends.

### G. *The missing filter*

Those metrics are not in dispute. The question is different. Takahashi measures learned structure and control capacity per joule. Fagan measures causal or goal-directed *work* per unit irreversible information — his numerator is outward causal efficacy, not inward model quality. Perrier measures energy efficiency relative to a behavioural intelligence score. Each metric is physically meaningful. None, however, makes *adaptively useful internal structure itself* the explicit numerator. The unresolved definitional question is therefore not whether intelligence is physically costly. It is which internal distinctions count as intelligence-relevant rather than merely ordered or merely benchmark-effective.

The usefulness filter belongs at the centre. Kolchinsky and Wolpert [13] have already formalised semantic information as the correlations causally necessary for maintaining existence, and the information bottleneck [14] already optimises relevance-filtered compression. The contribution here is not the filter itself but the *combination*: a relevance-filtered structure term ($C_u$), a thermodynamic acquisition-efficiency term ($I_{\text{eff}}$), and a behavioural axis ($A$) evaluated jointly rather than collapsed to a single scalar.

## IV. Why useful structure, not mere order

Open systems can create and maintain local order by drawing on free-energy gradients and exporting entropy. That fact alone does not imply intelligence. Self-organisation is broader than intelligence because many systems can stabilise, amplify, or pattern themselves without forming internal distinctions that improve their future coupling to the world.

To count as intelligent, a system must therefore do more than resist disorder. It must form or maintain internal organisation that improves prediction, control, recovery, or preservation of viable states. The relevant contrast is not between order and disorder, but between order that matters and order that does not.

Prediction is necessary but not sufficient. A system that perfectly tracks irrelevant fluctuations is informative but not intelligent. Control is necessary but not sufficient either. A system can exert large causal impact while remaining brittle, myopic, or destructive to its own viability. Intelligent organisation sits at the intersection: it selectively retains distinctions that improve future adaptive performance under resource constraint.

Recursive self-modelling, long-horizon planning, and reflective policy revision are important amplifiers of intelligence, but they are not the minimal criterion. The minimal criterion is usefulness: internal structure that is world-sensitive, action-guiding, and improves adaptive success.

## V. Proposed definition

**Intelligence** is the resource-bounded capacity of an adaptive system to acquire, compress, and deploy useful internal structure so as to improve prediction, control, and robustness across a range of environments.

The definition is explicitly resource-bounded: intelligence is never abstracted away from finite energy, time, memory, or dissipation. It is also explicitly selective. A large random table, a brittle policy, or a merely ornate dynamical system is not intelligent simply because it contains many states. What matters is whether the retained structure helps the system track relevant regularities, choose better actions, preserve viable futures, or recover from perturbation.

The term "model-based" is used here in a functional rather than a narrowly symbolic sense. The required internal structure may be explicit or implicit, learned or evolved, centralised or distributed across a collective. What matters is not the representational format but whether the system carries action-guiding distinctions about the world that improve adaptive performance.

## VI. Proposal: capability, useful structure, and acquisition efficiency

Intelligence requires evaluation along at least two axes. A system can be highly capable but inefficient, or efficient but narrow. Two evaluative axes are therefore distinguished—capability and efficiency—and one mediating state variable, useful internal structure.

### A. Capability: adaptive reach

Let $\mu$ denote a reference distribution over environments or task families, and let $S_{e(\pi)} \in [0, 1]$ denote the probability that system $\pi$ maintains the relevant viability condition or achieves the relevant goal in environment $e$. Define adaptive reach as

$$A(\pi; \mu) = E_{e\sim\mu}\left[S_{e(\pi)}\right] \tag{1}$$

This definition can be specialised in several ways: average-case, worst-case, thresholded, or transfer-focused. Its purpose is simple. Adaptive reach quantifies how broadly a system can deploy its organisation across varying conditions.

The viability or goal criterion is a parameter of the framework, not something derived from physics alone. For biological systems the criterion may concern survival or homeostatic integrity. For artificial systems it may concern externally specified goals or safety constraints. This normativity is unavoidable, but it is not unique to the present proposal. Behavioural definitions also depend on task distributions, reward structures, or priors. Here that dependence is made explicit.

### B. Useful internal structure

Let $M_t$ denote the system's internal state or internal model at time $t$. Let $H_t$ denote the system's past sensorimotor history up to but not including $t$; specifically, $H_t$ is the raw sensorimotor trajectory and does not include learned parameters, weights, or compressed summaries that constitute $M_t$. Let $Z^V_{t+\tau}$ denote those future environmental variables or latent states whose prediction or control matters for the relevant viability or goal criterion, evaluated at a horizon $\tau$ chosen to match the system's decision timescale. A concrete candidate for useful internal structure is then

$$C_u = I(M_t; Z^V_{t+\tau} \mid H_t) \tag{2}$$

This quantity is not all stored information. It is the predictive and control-relevant information that the system's internal organisation carries about the part of the world that matters for future adaptive success. The conditioning on $H_t$ is important: it ensures that $C_u$ measures what the agent's learned model contributes *beyond* the current episode's raw observations. For agents that have accumulated structure across prior experience, $M_t$ carries latent knowledge that $H_t$ alone does not contain, making the conditional mutual information non-trivially positive. In spirit, this treats $C_u$ as a semantic rather than merely syntactic information term: only correlations that matter for viability, prediction, or control count [13].

The compression term in the verbal definition matters here. An effective representation preserves information about $Z^V$ while discarding nuisance structure. This is the information-bottleneck intuition translated into an adaptive-control setting [14]. Different benchmarks may require different estimators, but the underlying quantity is always relevance-filtered internal structure, not raw state count or raw mutual information with arbitrary environmental variables.

### C. Efficiency: acquisition efficiency

Let $W_{\text{diss}}$ denote the physically meaningful work dissipated while the system learns, adapts, or reorganises. Define acquisition efficiency as

$$I_{\text{eff}} = \frac{\Delta C_u}{W_{\text{diss}}} \tag{3}$$

This is a flow variable. It measures how efficiently a system acquires additional useful structure over a learning or adaptation episode. It does *not* by itself quantify the intelligence of a fully trained system at steady state. If a deployed system is no longer learning, then $\Delta C_u = 0$ over that interval, which implies zero current acquisition efficiency, not zero intelligence. Deployed intelligence is captured by the stock variable $C_u$ and by the behavioural axis $A$.

Cross-substrate comparisons at different temperatures require a temperature-normalised form. If $C_u$ is measured in bits, one natural choice is

$$I^*_{\text{eff}} = \frac{k_B \bar{T} \ln 2 \; \Delta C_u}{W_{\text{diss}}} \tag{4}$$

which is dimensionless and makes the relation to Landauer-scale bounds explicit. Within a fixed hardware class and temperature regime, bits per joule remain a convenient reporting unit.

*D. What Still, empowerment, and active inference each contribute*

Still and colleagues support the predictive side of the proposal: nonpredictive information stored by a driven system provides a stepwise lower bound on dissipated work, $\langle W_{\text{diss}}\rangle \geq k_B T I_{\text{nonpred}}$ [10]. That inequality is an anchor for the claim that arbitrary stored complexity should not appear in the numerator; it does not, however, set an equality for finite protocols with additional irreversibility sources. It does *not* by itself derive intelligent control. Control enters through agency-sensitive quantities such as empowerment [15] and through closed-loop frameworks such as active inference [5], [7]. Prediction and control are therefore treated as complementary requirements. Still supports the filter against useless memory; empowerment and active inference support the requirement that retained structure must also be deployable in action.

*E. Relation to the earlier exploratory equation*

A natural but imprecise formulation would use $\Delta\frac{C}{E-T\Delta S}$. That expression captures the intuition that intelligence involves complexity, energy, and entropy, but it leaves the numerator semantically loose and introduces ambiguity about units. The formulation above replaces raw complexity with usefulness-filtered structure and treats the energetic term directly as dissipated work.

## VII. Illustrative toy example

Consider a minimal control task. A binary latent variable $z \in \{0, 1\}$ determines which action keeps an agent inside the viability set. Observations also contain a nuisance bit $n \in \{0, 1\}$ that is irrelevant to viability. Three agents are trained on the same hardware and task family over $K = 8$ i.i.d. episodes, each presenting a uniformly drawn $(z, n)$ pair with reward $r = \mathbf{1}\{a = z\}$.

The first agent learns the latent rule: its internal representation is $M = z$, compressing the two-bit observation to the one bit that matters. The second memorises all four observation–action pairs, storing $M = (z, n)$. The third memorises only the nuisance bit, setting $M = n$.

Because each agent faces a fresh i.i.d. draw with no prior episode history, $H_t = \emptyset$ and the general form $C_u = I(M_t; Z^V_{t+\tau} \mid H_t)$ reduces to $C_u = I(M; Z)$. This special case demonstrates the *compression* aspect of useful structure — the ability to extract $Z$ from a noisy observation — but does not exercise the *predictive* aspect that the conditioning on $H_t$ captures in sequential settings. We compute $C_u$ exactly by constructing the joint distribution $P(M, Z)$ for each agent, where $Z = z$ is the viability-relevant variable. Because $z$ and $n$ are independent and uniform, the joint distributions are:

- **Rule learner:** $P(M = i, Z = j) = 1/2$ when $i = j$ and zero otherwise. This gives $I(M; Z) = H(Z) = 1$ bit.
- **Lookup table:** $P(M = (i, k), Z = j) = 1/4$ when $i = j$ and zero otherwise. Again $I(M; Z) = 1$ bit, but $H(M) = 2$ bits — one bit of nuisance is stored without increasing useful structure.
- **Noise memoriser:** $P(M = k, Z = j) = 1/4$ for all $k, j$, so $I(M; Z) = 0$.

TABLE I

Exact toy calculation. $C_u$ is computed analytically from joint distributions. $W_{\text{diss}}$ is counted at the Landauer floor ($k_B T \ln 2$ per bit erasure, $T = 300$ K). Adaptive reach $A$ is estimated over 1000 environments with varying nuisance dimensionality ($n_{\max} \in \{1, 2, 3, 4\}$, viability threshold $\theta = 0.8$). Code: compute_cu.py.

| Agent | $H(M)$ | $C_u$ | $W_{\text{diss}}$ | $I^*_{\text{eff}}$ | $A$ | Accuracy (train) |
|---|---|---|---|---|---|---|
| Latent-rule learner | 1 bit | 1 bit | 16 $k_B T \ln 2$ | 0.0625 | 1.00 | 1.00 |
| Lookup-table memoriser | 2 bits | 1 bit | 24 $k_B T \ln 2$ | 0.0417 | 0.54 | 1.00 |
| Noise memoriser | 1 bit | 0 bits | 16 $k_B T \ln 2$ | 0.0000 | 0.00 | 0.50 |

Dissipated work is estimated at the Landauer floor. Each training episode requires a fixed number of irreversible bit erasures determined by the agent's architecture: two erasures per episode for a one-bit register (compare and update), three for a four-entry lookup table (address, compare, update). Over $K = 8$ episodes, the rule learner and noise memoriser each dissipate $W_{\text{diss}} = 16 k_B T \ln 2$ while the lookup table dissipates $W_{\text{diss}} = 24 k_B T \ln 2$. At $T = 300$ K, one $k_B T \ln 2 \approx 2.87$ zJ.

Several observations follow. First, $C_u$ isolates the useful information in this toy: both the rule learner and the lookup table carry one bit about $Z$, but the lookup table wastes an additional bit on nuisance structure. A raw information metric would credit the lookup table for storing more total structure. Second, the rule learner is 50% more efficient ($I^*_{\text{eff}} = 0.0625$ vs. 0.0417) because it avoids the addressing overhead of a larger memory. Third, all three agents can be tied or separated on training-distribution accuracy, but only $C_u$ and $A$ jointly distinguish the rule learner as the agent with genuinely useful structure. Fourth, the noise memoriser stores one bit and dissipates the same energy as the rule learner, yet $C_u = 0$ — it has learned nothing about viability. The framework captures this cleanly: useful structure is not raw stored information but relevance-filtered information about $Z^V$. Viewed through the information bottleneck, the rule learner implements the IB-optimal compression for this channel: it sits at the knee of the IB curve, achieving maximum $I(M; Z)$ at minimum encoding rate $I(M; O)$. The lookup table achieves the same $I(M; Z)$ at double the rate, placing it strictly inside the IB-achievable region [14].

## VIII. Empirical programme and falsifiable tests

Three comparative tests are central.

*A. Test 1: usefulness-filtered structure should outpredict size-based proxies*

Within a fixed task family and hardware class, estimates of $C_u$ should predict held-out adaptation, transfer, and robustness better than raw parameter count, raw energy use, or raw learned bits. A purely reactive controller may achieve viability within a narrow niche, but it will score low on adaptive reach across the environment distribution $\mu$; the framework captures this through the separation of $A$ from $C_u$. In machine learning, this test can be conducted by training multiple architectures on the same hardware, measuring wall-plug or board-level energy, estimating $C_u$ with a task-relative proxy, and then testing out-of-distribution adaptation.

*B. Test 2: nuisance memory should raise raw information but not intelligence*

Interventions that increase stored structure without increasing information about $Z^V$ should raise raw complexity measures more than they raise $C_u$, $I_{\text{eff}}$, or $A$. This is the core negative test against the claim that any entropy-resisting or information-rich process should count as intelligent.

*C. Test 3: control-improving interventions should matter more than scale alone*

Interventions that improve representation quality, planning depth, or controllability should improve adaptive reach more reliably than interventions that merely increase scale while leaving relevance-filtered structure unchanged.

*D. Candidate validation domains*

**Machine learning.** Estimate $C_u$ using viability-relevant latent variables when available, or task-relative predictive information and controllability proxies when they are not. Compare $C_u$, $I_{\text{eff}}$, and $A$ against size, accuracy, and energy-only baselines. The Stanford Intelligence-per-Watt benchmark provides a useful empirical template, though it does not yet estimate useful structure directly [23].

**Biological agents.** Compare organisms or embodied robots on context switching, prediction under perturbation, and recovery from shocks while normalising against metabolic or device-level energy expenditure.

**Collective systems.** Estimate distributed useful structure through coordination quality, error correction, and resilience under shocks relative to communication and energetic cost.

The framework weakens if raw size or raw learned bits predict adaptive success better than usefulness-filtered structure, or if systems that accumulate clearly irrelevant structure score equally well under the proposed quantities.

## IX. Boundaries and limitations

The central claim is conceptual and operational: a thermodynamic definition of intelligence should include a usefulness filter on internal structure.

Several limits should be stated plainly. First, the measurement problem remains open. Equation (2) is a concrete candidate, not a finished universal estimator. Some domains will require latent-state models; others will need predictive-control proxies. Second, the proposal is not a complete scalar ranking of systems. The relevant object is an intelligence profile: adaptive reach $A$, useful-structure stock $C_u$, and acquisition efficiency $I_{\text{eff}}$. Third, active inference already provides a rich account of viability and attracting states; the present contribution is narrower, namely an explicit efficiency term for the acquisition of relevance-filtered structure. Fourth, the framework does not require explicit symbolic world models. Implicit, embodied, and distributed representations can all count if they carry the right relevance-filtered distinctions. Fifth, when dissipated work approaches zero (near-reversible computation), $I_{\text{eff}}$ diverges; in practice the Landauer limit $k_B T \ln 2$ per bit erasure provides a natural floor and the quantity remains well-defined for any physically realisable learner. Sixth, when viability is multi-dimensional, $C_u$ should be computed as the joint mutual information $I\big(M_t; Z_1^V, \ldots, Z_k^V \mid H_t\big)$; sums of per-criterion marginals can undercount synergistic structure.

The framework succeeds if it makes the target of measurement clearer than existing alternatives and generates discriminating empirical tests.

## X. Conclusion

Recent thermodynamic formalisms have shown that learning, computation, and control admit physically meaningful efficiency measures. Those measures are necessary but not yet sufficient as definitions of intelligence unless they distinguish useful internal structure from structure that is merely abundant or causally potent.

On the account proposed here, intelligence is the resource-bounded capacity of an adaptive system to acquire, compress, and deploy useful internal structure so as to improve prediction, control, and robustness across environments. Adaptive reach captures the breadth of deployment. Acquisition efficiency captures the energetic efficiency with which useful structure is gained. The key unresolved task is to operationalise useful structure well enough that it predicts adaptation better than size-based and energy-only proxies.

If that task succeeds, the field gains more than another bits-per-joule curve. It gains a principled way to separate intelligent order from merely persistent order, and therefore a more credible thermodynamic science of intelligence.

## XI. Author contributions

B.D. conceived the framework, conducted the literature review, performed the computations, and wrote the manuscript.

## XII. Funding

This research received no external funding.

## XIII. Data availability statement

The computation code used to generate Table I and the neural-network validation suite will be made publicly available at github.com/199-biotechnologies/fti upon publication.

TABLE II
POSITIONING RELATIVE TO ADJACENT TRADITIONS AND RECENT FORMALISMS

| **Tradition / formalism** | **What it contributes** | **What this proposal adds** |
|---|---|---|
| Behavioural / psychometric [1], [2] | Adaptive success across environments | Physical cost and an explicit account of useful structure |
| Active inference / empowerment [5], [15] | Internal models, control, future options, viability | A thermodynamic acquisition-efficiency term |
| Thermodynamics of prediction [10] | Why nonpredictive memory is dissipatively wasteful | Extension from predictive usefulness to predictive *and* control relevance |
| Epiplexity / empowerment per joule [18] | Learned structure and control capacity per unit energy | A usefulness filter on which learned structure should count |
| Conservation-Congruent Encoding [19] | Causal or goal-directed work per unit irreversible information | Shifts the numerator from work done to useful internal structure |
| Watts-per-intelligence [20] | Energy efficiency relative to a behavioural score | Separates benchmark performance from relevance-filtered internal structure |
| Semantic information / information bottleneck [13], [14] | Formal relevance and selective compression | Places that relevance criterion inside a thermodynamic intelligence proposal |

TABLE III
EARLY VALIDATION PROGRAMME

| **Domain** | **Candidate estimate of $C_u$** | **Cost term** | **What would support the theory?** |
|---|---|---|---|
| Machine learning | Task-relative predictive/control information about viability-relevant latents | Wall-plug or board-level joules | Better prediction of transfer and robustness than size or raw energy |
| Biological agents | Prediction- and control-relevant behavioural state information | Metabolic or neural expenditure | Broader adaptive reach per unit dissipated work |
| Collective systems | Distributed coordination information relevant to shared viability | Communication and energetic cost | More resilient coordination without credit for irrelevant internal complexity |

The scripts include `compute_cu.py` (exact toy calculation), `neural_experiment.py` (Level 2 experiment), `validation_suite.py` (7-agent cross-validation), and `hypothesis_battery.py` (40-hypothesis stress test).

## XIV. CONFLICTS OF INTEREST

The author is the founder of 199 Biotechnologies (SG) Pte Ltd, which develops AI infrastructure tools. The research was conducted independently and the company had no role in the design or writing of this work.